# REDUCED 30% SCANNING TIME 3D MULTIPLEXER INTEGRATED CIRCUIT APPLIED TO LARGE ARRAY FORMAT 20KHZ FREQUENCY INKJET PRINT HEADS

*Jian-Chiun Liou[1], and Fan-Gang Tseng[1,2,3]*

[1]MEMS Inst., [2]Engineering and System Science Dept.
National Tsing Hua University, Taiwan, ROC.
[3]Division of Mechanics, Research Center for Applied Sciences, Academia Sinica, Taiwan, ROC

**ABSTRACT**

Enhancement of the number and array density of nozzles within an inkjet head chip is one of the keys to raise the printing speed and printing resolutions. However, traditional 2D architecture of driving circuits can not meet the requirement for high scanning speed and low data accessing points when nozzle numbers greater than 1000. This paper proposes a novel architecture of high-selection-speed three-dimensional data registration for inkjet applications. With the configuration of three-dimensional data registration, the number of data accessing points as well as the scanning lines can be greatly reduced for large array inkjet printheads with nozzles numbering more than 1000. This IC (Integrated Circuit) architecture involves three-dimensional multiplexing with the provision of a gating transistor for each ink firing resistor, where ink firing resistors are triggered only by the selection of their associated gating transistors. Three signals: selection (S), address (A), and power supply (P), are employed together to activate a nozzle for droplet ejection. The smart printhead controller has been designed by a 0.35 um CMOS process with a total circuit area, 2500 ×2500 μm2, which is 80% of the ciruciut area by 2D configuration for 1000 nozzles. Experiment results demonstrate the functionality of the fabricated IC in operation, signal transmission and a potential to control more than 1000 nozzles with only 31 data access points and reduced 30% scanning time.

## 1. INTRODUCTION

One of the major goals for advanced inkjet print head design is to optimize printing quality and speed while minimizing cost. To achieve this, the most commonly employed architecture of driving IC for commercial inkjet print head nowadays is carried out by two-dimensional arrayed-switches. The data accessing points will be X(Pads) $=2\times \sqrt[2]{Y}$ +1 (Y~Nozzles), which is equal to 21 if the nozzle number is 100. However, if the number of nozzles increases to thousands in a large array format inkjet print head, not only the data accessing points will be easily increased to hundreds, but also the scanning time will significantly rise, which deteriorates the performance of inkjet printing[1][2].

In this study, all of the jets of the print head are controlled by very few input lines：a pulse shape (ENABLE), a data line (DATA), a bit shift clock (BIT SHFT), a state clearing pulse, 5-volt supply for the logic devices, a higher voltage for energizing the heater resisters, and a ground line .Figure 1 is a block diagram showing an internal configuration of the digital driver provided in the inkjet print head device. A three dimensional data registration is schemed to reduce the number of data accessing points as well as scanning lines for large array inkjet print heads with over 1000 nozzles. The total numbers of data accessing points will be X=$3\times \sqrt[3]{Y}$ +1, which is 31 for 1000 nozzles by the 3D novel design, a dramatic reduction from 68 if operated by the traditional 2D scheme. The property comparison among 1D, 2D, and 3D architectures is listed in Table 1. As the nozzle number increases, a higher order circuit can effectively reduce the pad number.

**Table 1 Property comparison among 1D, 2D, and 3D driving schemes**

| X:Pads、Y:Nozzles | X~Y+1 | X=$2\times \sqrt[2]{Y}$ +1 | X=$3\times \sqrt[3]{Y}$ +1 |
|---|---|---|---|
| Nozzles | 1000 | 1024 | 1000 |
| Heaters | 1000 | 1024 | 1000 |
| Resolution (dpi) | 300 | 300~600 | >600 |
| Print Swath(in) | 1/6 | 1/3 | >1/3 |
| Interconnect Pad | 1001 | 65 | 31 |

Several options were considered for a conventional two-dimensional addressing circuitry. The three most promising were row-column de-multiplexing (demux), blocking diodes, and a proprietary passive enhanced multiplexing scheme. The row-column scheme was chosen. This scheme reduces the number of pads from $X(Pads) = 2 \times \sqrt[2]{Y} + 1$ for a print head with Y resistors. When the row and column connected with the resistor cell are applied a positive voltage, the transistor will conduct and generate a current flow through the resistor. Then the resistor will heat the ink, generating bubbles to spray the ink droplets.

## 2. DESIGN

In the proposed novel 3D design, different from the 2D one, as shown in Figure 1, the digital driver includes a clock-control circuit, a serial/parallel-conversion circuit, a latch circuit, a level shifter, a D/A converter comprised of a decoder, and an output buffer comprised of an operation amplifier.

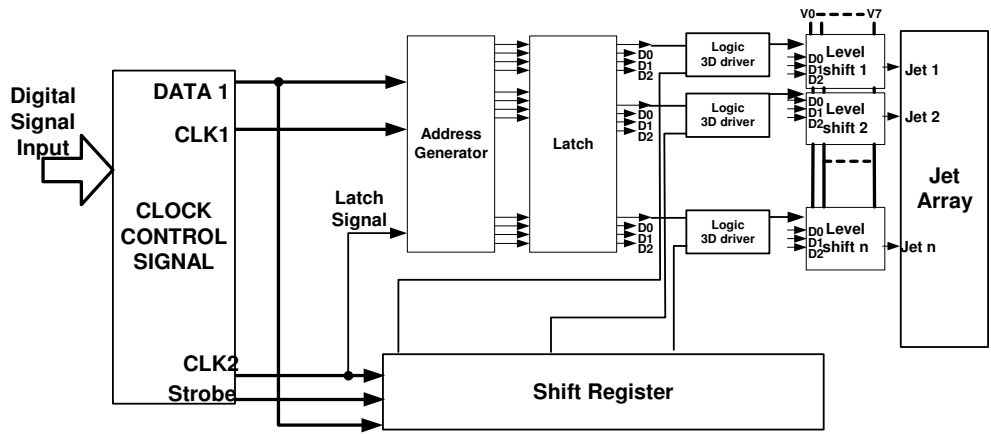

**Figure 1: An inkjet print head digital driver block diagram**

The D/A converter receives a gray-scale reference voltage from an external source. The clock-control circuit receives control signals from an external control circuit. Based on the received control signals, the clock-control circuit attends to control of the latch circuit, the D/A converter, and the output buffer by using a latch-control signal. The received inkjet data needs to be converted into data at an optimal transfer rate (frequency) in order to conform to the inkjet characteristics. To this end, the clock-control circuit divides the 8-bit inkjet signals supplied to the data driver, with an aim of lowing the operation frequency. The serial/parallel-conversion circuit converts serial signals of a plurality of channels into parallel signals, and supplies the parallel signals to the latch circuit. The latch circuit temporarily stores therein the received parallel signals, and supplies same to the level shifter and the D/A converter at predetermined timings. The level shifter converts a logic level ranging approximately from 3.5 V to 5 V into a inkjet voltage level that ranges from 7.5V to 8.5V for various heater resistors as a result of variation processing conditions. The control circuit is divided into two parts: the pass-gate device (for signal path) and the power device (for power path), as shown in Figure 2. The pass gate device is controlled by A selection and S selection, while the power line is controlled by P selection. To activate one heater, all P, A, and S selections are required to be turned on at once. For example, to turn on the heater 1, P1, A1 and S1 need to be set high at the same time. The scanning time is reduced up to 30% (The scanning speed is also increased by 3 times) thanks to the great reduction of lines for 3D scanning, rather than 2D scanning.

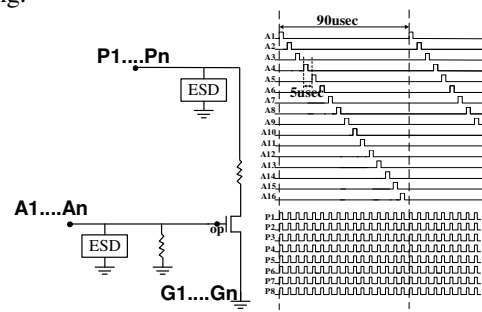

**(A) 2D method**

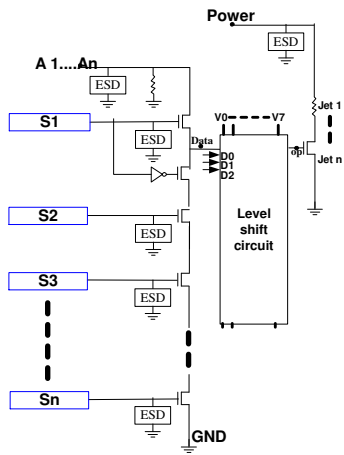

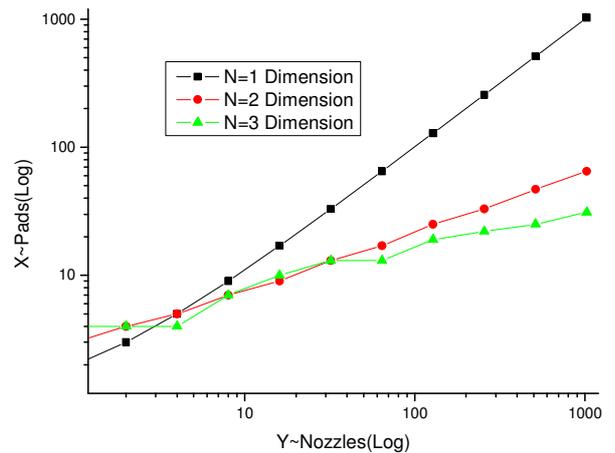

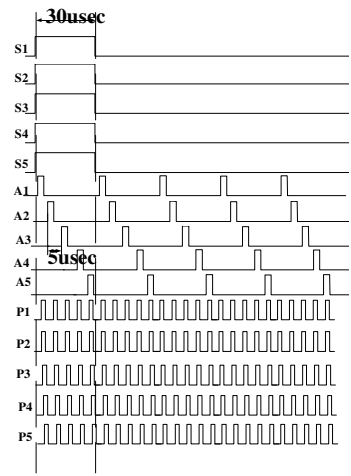

**(B) 3D method**

**Figure 2**：**The sequence of driving signals for (A) 2D method and (B) 3D method and driving scheme**

Figure 3 shows the numbers of the required connection pads for 1D, 2D and 3D control circuit. As the nozzle number increases, the required connection pads are increasing but with different rates. The 1D case increases most rapidly while the 3D one increases at the slowest pace. There are two important intersections in the three curves, which are, the one with the nozzle number of 10 and the other, 30. When the nozzle number is much larger than 100, the 3D architecture can dramatically reduce the pad number in tens.

**Figure 3**：**The numbers of required connection pads for 1D, 2D and 3D control circuit**

### 3. ANALYSIS AND RESULTS

Figure 4 is the input-output relationship of the ASIC we observe with the Logic Analysis equipment. The output signal is matched to the designed function .An input signals waveform concept can be used to TIJ to address each jet . The depicted integrated circuit TIJ transducer array serves 1000 jets and includes data interfacing , jet addressing. The desired signal for S selections and A selections can be pre-registered and latched in the circuit for one time writing. Combining this scheme with the P selection control, simultaneous 3D driving can be realized. The enhancement will be even sounder in larger nozzle arrays.

Figure 5 illustrates the transient simulation of the input and output signal of the level shift device[3][4][5]. The signal demonstrates that not only the switch speed is higher by the level shift device than that of one without level shift circuit , but also the voltage has been enhanced to 5V [6][7].An adjustable voltage pulse from 7.98V to 8.02V is applied to the various heater resistors thanks to the processing condition.

The characteristic of pass-gate device with 0.35 μm CMOS technology measured by HP 4156 Semiconductor Parameter Analyzer show that the break-down voltage is close to 9 volts, higher enough for A selection signal to pass through, and illustrate that a driver functional circuit with a break-down voltage of 9 volts and gate threshold voltage of 1 volt, respectively.

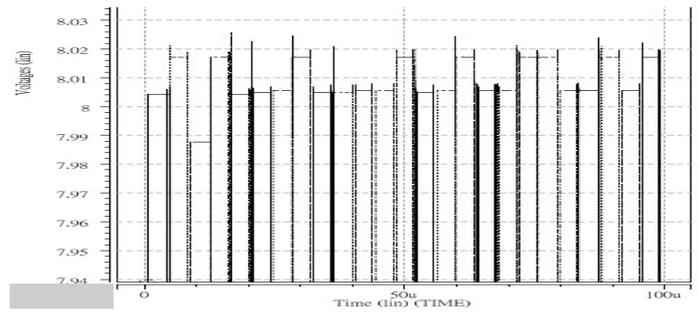

**Figure 5**：Transient simulation of the input and output signals of the level shift device

Figure 6 shows a photograph of the realized prototype. Much care must be taken in the layout of the metal layers in order to avoid electro-migration. The technology used was a two-poly four-metal (2P4M) 0.35um twin-well CMOS technology. The visualization of the droplet ejection at a frequency over 20kHz has been performed with pseudo-cinematography. This method is based on the stroboscopic principle. A short heating pulse drives a micro heating element and heats a thin liquid layer on the heater in less than 3usec up to more than 300°C . At a normal pressure of 1 bar water reaches the thermodynamic limit of superheating , at a temperature of approximately 312°C .

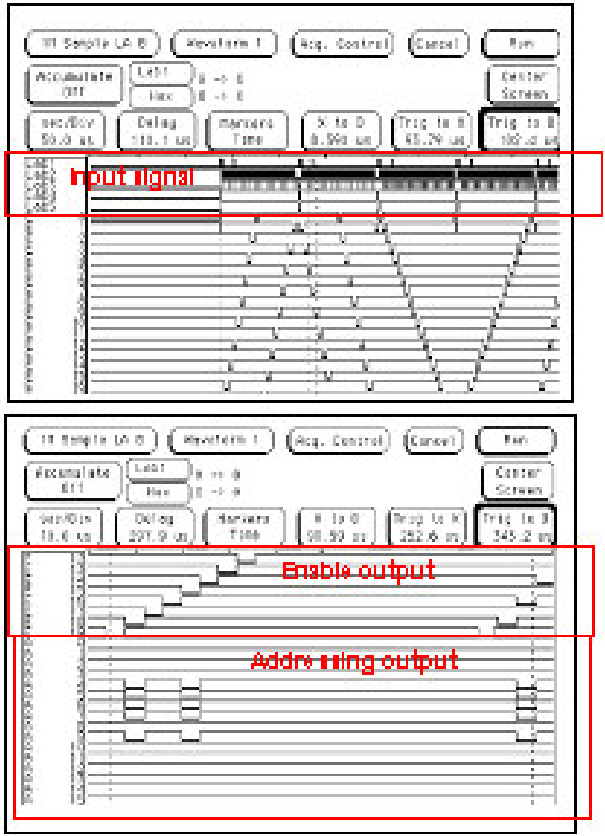

**Figure 4. The Logic Analysis signal**

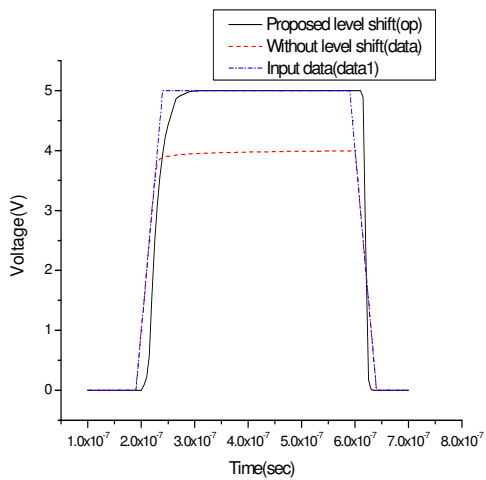

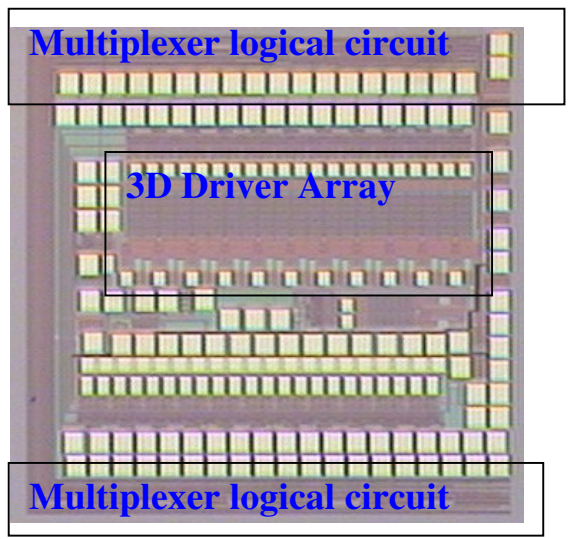

**Figure 6**：Chip photograph

The testing result of a 2D IC, as shown in Figure 7, demonstrats that for firing the last nozzle (number 125),

the firing singals need to wait for the scanning from A1 to A16 and P1 to P8, taking significant time.

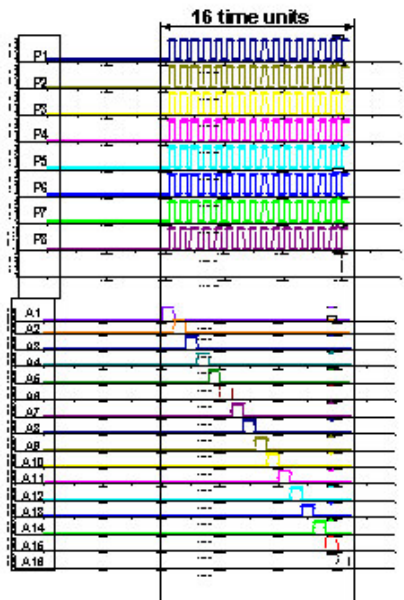

**Figure 7. 2D signal results**

However, the same task can be finished in a much shorter time in a 3D IC, as shown in Figure 8. It takes only the scanning from A1 to A5 and P1 to P5 to fire the last (number 125) nozzle. The time reduction is about 70%, providing much faster driving signals for a high density and large array inkjet printhead.Figure 9 Visualization of droplet ejection.

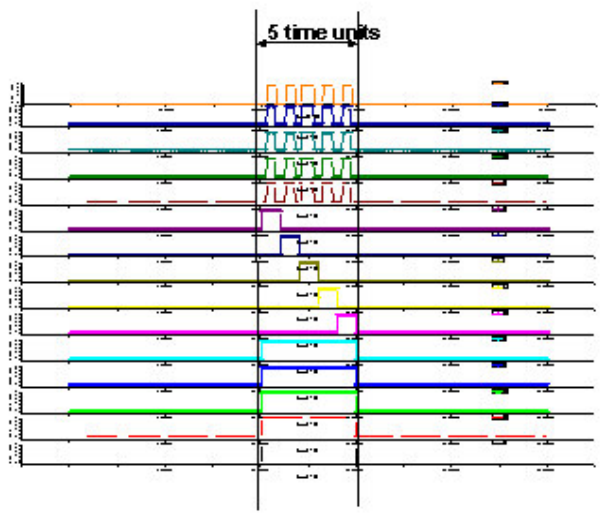

**Figure 8. 3D signal results**

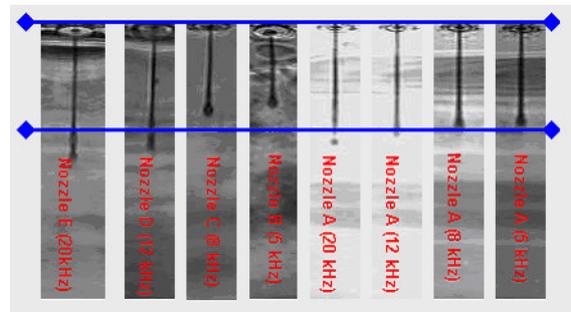

**Figure 9  Visualization of droplet ejection**

## 4.CONCLUSIONS

This paper proposes a novel architecture of high-selection-speed 3D-data-registration for driving large-array inkjet printhead. The 3D driving architecture has successfully reduced the total numbers of control pads into 31 for 1000 nozzles as well as the scanning time up to 30% with a higher signal rising speed and smaller circuit area. All the sub-circuits, including power control, digital I/O, analog-to-digital converter, power drivers were integrated into a single device. This circuit has been designed, fabricated, and characterized.It demonstrated not only the functionality in application but also the consistency between simulation and experiment results.

## 6. AUTHOR BIOGRAPHY

Jian-Chiun Liou was born in 1974 in Kaoshiung, Taiwan, ROC. He received his M.S. degree from the Institute of Electronic of National Taiwan Ocean University, Taiwan, and currently a Ph.D. student with The MEMS Institute of National Tsing Hua U. .He joined the Printing technology Development and Manufacturing Section of OES/ITRI in 1999 . His research interest is in the ASIC design, MEMS technology and integration of inkjet printhead processes .

Fan-Gang Tseng received the B.S. degree in power mechanical engineering from National Tsing Hua University, Taiwan in 1989, and the M.S. degree from the Institute of Applied Mechanics in National Taiwan University, Taiwan, in 1991. In 1998, he received his Ph.D. degree in mechanical engineering from the University of California, Los Angeles, USA (UCLA). After one year with USC/Information Science Institute as a senior engineer working on a new microfabrication process, EFAB, he became an assistant professor with Engineering and System Science Department of National Tsing-Hua University, Taiwan from August, 1999, and advanced to professor in August, 2006. His research interests are in the fields of Bio-MEMS/Nano and Nano/Micro-Fluidic Systems. He received 19 patents, wrote book chapters of "Micro Droplet Generators" in MEMS Handbook by CRC press and "Technological Aspects of Protein Microarrays and Nanoarrays" in Protein Microarrays by Jones and Bartlett Publishers, published more than 50 SCI Journal papers and 120 conference technical papers in MEMS, Bio-N/MEMS, and micro/nano fluidics related fields, and co-chaired in many technique sessions including IS$^3$M, H.K. in 2000, and IEEE Transducers'01, Munich, Germany in 2001. He received several awards, including Mr. Wu, Da-Yo Memorial Award from National Science Council, Taiwan (2005), three best paper/poster awards (1991, 2003, 2004, and 2005), NTHU new faculty research award (2002), NTHU outstanding teaching award (2002), NTHU academic booster award (2001), and NSC research award (2000).